\documentstyle{article}
\newtheorem{theorem}{Theorem}

\begin{document}
\normalsize
\begin{center}
{\bf There are Asymmetric Minimizers for the 
One-Dimensional Ginzburg-Landau Model of Superconductivity } \\
by \\
Stuart P. Hastings\\
And\\
William C. Troy\\
Department of Mathematics\\
University of Pittsburgh\\
Pittsburgh, PA  15260
\end{center}
\medskip
\medskip
\begin{abstract}  We study a boundary value problem associated with a
system of two second order differential equations with cubic nonlinearity
which model a film of superconductor material subjected to a tangential
magnetic field.  We show that for an appropriate range of parameters
there are {\it asymmetric} solutions, and only trivial {\it symmetric}
solutions.  We then correct an error of the authors in [9] and show
that the associated energy function is negative for the asymmetric
solutions, and zero for the trivial symmetric solution.  It
follows that a global minimizer of the energy is asymmetric.  This
property resolves a conjecture of Marcus [13].
\end{abstract}
\bigskip

\bigskip

\bigskip

\medskip
\section{  Introduction}
\medskip
\hspace{\parindent}
 
In this paper we continue our recent studies ([9], [10]) of 
the one dimensional Ginzburg-Landau model [8] for 
superconductors.  Our main objective is to investigate 
the model for the existence of asymmetric minimizers of 
the appropriate energy integral.  It is expected that the 
physically interesting solutions will be energy 
minimizers.  These solutions satisfy a symmetric 
boundary value problem which was known to have a set 
of symmetric solutions.  We will show, however, that in 
one space dimension, for some parameter values
 the energy minimizer is an asymmetric solution 
to this symmetric boundary value problem. 

   The problem may be compared to studies in two 
dimensions, where the symmetry of the energy 
minimizer is unresolved.  Recent papers on this include [12] and 
[14].  Work in two dimensions has largely dealt with a 
reduced problem in which variations in the magnetic 
field are ignored.  In contrast, it is possible to include 
the magnetic field as a variable in the analysis of the 
one dimensional model. 

In order to properly describe our results we first need 
to give a brief development of the problem as well as a 
summary of our previous investigations.  

In 1950 Ginzburg and Landau [8] proposed a model for the
electromagnetic properties of a film of
superconducting material of width $d$ subjected to a tangential external
magnetic field.  Under the assumption that all quantities are functions
only of the transverse coordinate, they proposed that the electromagnetic
properties of the superconducting material are described by a pair
$(\tilde \phi, \tilde a)$ which minimizes the free energy functional
$${\cal G} = \frac 1d \int^{d/2}_{-d/2}\Big(\tilde{\phi}^2(\tilde{\phi}^2-2)
+ \frac{2(\tilde{\phi}')^2}{k^2} + 2 \tilde{\phi}^2\tilde{a}^2 + 2(\tilde
{a}'- h_e)^2\Big)dx\eqno(1.1)$$
The functional $\cal G$ is now known as the Ginzburg-Landau energy and
provides a measure of the difference between normal and superconducting
states of the material.  The variable $\tilde \phi$ is the order
parameter which measures the density of superconducting electrons, and
$\tilde a$ is the magnetic field potential.  Also, $h_e$ is the external
magnetic field, and $k$ is the dimensionless material constant
distinguishing different superconductors, i.e. $0 < k < \frac{1}{\sqrt
2}$ for type I superconductors, and $k > \frac{1}{\sqrt 2}$ for type II
superconductors (see also [7]).  The minimizer requirement,
that $\cal G$ be stationary with respect to general first order
variations of the functions $\tilde \phi$ and $\tilde a$, leads to the
boundary value problem
$$\tilde{\phi}^{''} = k^2 \tilde{\phi} (\tilde{\phi}^2 + \tilde{a}^2-1),
\eqno (1.2)$$
$$\tilde{a}^{''} = \tilde{\phi}^2\tilde{a}, \eqno(1.3)$$
$$\tilde{\phi}'\Big(\pm\frac d2\big)=0,\,\,\tilde {a}'\Big(\pm\frac d
2\big)=h_e.\eqno(1.4)$$
 
It is routine to prove that ${\cal G}$ has a smooth minimizer 
satisfying (1.2) - (1.4) for any positive $h_e.$  In 1964 
Marcus [13] investigated the problem (1.2)-(1.4) and gave 
arguments which imply that a non-trivial minimizer of ${\cal G}$ 
should also satisfy 

$$\tilde{\phi }(x)>0\quad {\rm f}{\rm o}{\rm r}\ \ {\rm a}{\rm l}
{\rm l}\quad x\in\Big[-\frac d2,\frac d2\Big],\eqno(1.5)$$
and therefore this is the only kind of solution we consider.
A solution of (1.2)-(1.3) is called symmetric if
$$\tilde{\phi}'(0) = 0 \quad {\rm and}\quad \tilde{a}(0) = 0.\eqno(1.6)$$
It follows from (1.2) and (1.3) that if $\tilde{\phi}'(0) = \tilde{a}(0) =
0,$ then $\tilde{\phi}$ is an even function and $\tilde{a}$ is odd. 
Thus $\tilde{\phi}$ is symmetric with respect to the origin and
$\tilde{a}$ is
antisymmetric.  If (1.6)
does not hold then the solution is called asymmetric.  Marcus makes the
conjecture that a minimizer of $\cal G$ is probably a symmetric solution
satisfying (1.2)-(1.6).  However, he leaves open the possibility that asymmetric
solutions may also exist.

In later work Odeh [15] gave criteria for asymmetric solutions to
exist by bifurcation as $h_e$ increases, and in [2]-[5], Bolley and 
Helffer
give results implying that these criteria
are satisfied for each $k>0.$  The existence of at least one
symmetric solution has been investigated by Odeh [15], Wang and Yang
[18], Yang [19], and also Bolley and Helffer ([2]-[5]).  Numerical
studies, such as the work of Seydel ([16], [17]), and also more recent
theoretical work of Kwong [11], predict that a range of parameters
exists for which asymmetric solutions and multiple symmetric solutions
coexist.  The work of Seydel [16] also predicts that there is a range of
parameters for which no non-trivial symmetric solutions exist, yet
asymmetric
solutions do exist.  Other papers, such as [6], considered the problem on 
an infinite interval, whereas in our work, the interval is large but
finite. None of these
studies addressed the physically important criterion of
whether the solutions of the problem (1.2)-(1.5) 
are actually global minimizers of the energy functional $\cal G.$
However, a recent paper by Aftalion [1] does discuss this problem, and includes
a conjecture that asymmetric solutions can have a lower energy than
the symmetric solutions.  This is confirmed in the current paper.

In two recent papers ([9], [10]) we began our investigation of the
problem (1.2)-(1.5) with the goal of proving the existence of solutions
predicted by the numerical studies described above.  First, we studied
the existence of multiple symmetric solutions and proved the following
result.

\begin{theorem}  (see [9])
\medskip
\hspace{\parindent}

{\bf(i)}  Let $k \in (0, \frac{1}{\sqrt 2}).$  If $d > 0$ is sufficiently
large there is a range of values of $h_e$ for which at least {\it
two} symmetric solutions exist.

{\bf(ii)}  Let $k > \frac{1}{\sqrt 2}.$  If $d > 0$ is sufficiently
large there is a range of values of $h_e$ for which at least {\it three}
symmetric solutions exist.
\end{theorem}

\medskip
\noindent
{\bf Remark:}
\hspace{\parindent}  In related studies, Bolley and Helffer ([2]-[5])
have investigated other properties of symmetric solutions,
including bifurcation analyses and the uniqueness of solutions.  Their
analysis assumes that $k$ tends to zero, whereas our results assume that
$k > 0$ is fixed and $d$ becomes large.

In [9] we shifted our attention from the symmetric 
solutions found in [10] to the study of asymmetric 
solutions.  As mentioned above, the numerical 
experiments of Seydel (in particular, see Fig.  6.10 in 
[16]) predict that there is a range of parameters in 
which there are no non-trivial symmetric solutions, yet 
asymmetric solutions do exist.  This leaves open the 
possibility that in this parameter range the minimizer of 
${\cal G}$ could be an asymmetric solution.  Thus, our goal in [9] 
was to prove that there is a range of parameters in 
which only asymmetric solutions exist, and that the 
energy ${\cal G}$ is minimized by such solutions.  The first step 
in proving this result is to find an upper bound on the 
values of $h_e$ for which a symmetric solution exists.  
Thus, for fixed $k>0$ and $d>0,$ we let $h_e^{*}$ denote the 
supremum of the set of all positive $h_e$ for which a 
{\em non-trivial\/} symmetric solution exists.  Since we are 
assuming that $d\gg 1$ we define $h_e^{sym}=\overline {\lim}_{d\to
\infty}h_e^{*}.$ 

\medskip
In [9]  we proved that if $k\ge\frac 1{\sqrt {2.01}},$ then $h_e^{
sym}<\sqrt {3}k$ for 
large $d$.
Unfortunately, a rescaling error led us to assert that this
inequality was sufficient to prove that there are 
asymmetric minimizers.  It turns out that we need a 
stronger estimate.  We shall show below that the 
inequality
$$h_e^{sym}<1.68k\eqno(1.7)$$
suffices.  This is obtained by a routine but tedious refinement of 
the proof in [9].  We shall not repeat this proof here.
However, in an appendix we describe the changes which 
must be made in [9] to obtain (1.7). 
We add that the proof of (1.7) is considerably easier for 
large $k.$ The proof for this case is in [9].
 However, we want to include values in the 
range of type I superconductors, and this is more 
difficult.  

\medskip
\noindent{\bf Statement of Main Results}
\medskip

In this paper we will make use of (1.7) in proving that ${\cal G}$ 
has an asymmetric minimizer.  We prove two main 
results.  First, we fix $k\ge\frac 1{\sqrt {2.01}}$ so that both type I and 
type II superconductors are included.  Then, in Theorem 
2, we consider large $d\gg 1$ and prove that there exists a 
family of small amplitude asymmetric solutions of the 
problem (1.2)-(1.5).  We will also prove that
 $\frac {h_e}k\ge 1.6831$ for large $d,$ for
each of the asymmetric solutions found in Theorem 2.  
This and (1.7) confirm the numerical prediction of Seydel 
in [16], that there is a parameter regime in which there 
are asymmetric solutions, and only trivial symmetric 
solutions.

\medskip

The work of Bolley and Helffer includes results which imply the existence
of asymmetric solutions.  The proofs, which must be pieced together from
several papers, are by bifurcation theory, and do not appear to give
the estimate of $h_e$ which we obtain and which is essential in our discussion
of whether these asymmetric solutions are energy minimizers.  
\medskip
The proof of Theorem 2 is given in Section 2.  
Next, in Theorem 3 (proved in Sections 3 and 4) we 
show that the energy is negative for the asymmetric 
solutions found in Theorem 2.  This will allow us to 
prove that for large $d,$ a global minimizer of ${\cal G}$ must 
be asymmetric. 
\medskip
    
In addition to showing that
asymmetric solutions exist and have small amplitude, we will prove that 
each of these has
exactly one critical point (a relative maximum) in the open interval
$(-\frac d2, \frac d2),$ and that the relative maximum occurs at a value
close to $-\frac d2.$  Because of these properties, we find it
convenient to rescale the problem.  We introduce parameters $r,
h, m$ and $M$, and a new independent variable $t$, by setting
$$r = \frac{1}{k^2}, \ h = \frac{h_e}{k}, \ m + M = kd, \ x =
\frac{d}{2(m + M)}(2t + m - M).\eqno(1.8)$$
Next, we define new dependent variables $\psi$ and $A$ by
$$\tilde{\phi}(x) = \beta \psi(t) \quad {\rm and} \quad \tilde{a}(x) = 
A(t),\eqno(1.9)$$
where $\beta=\tilde{\phi}(0)$. Then (1.1)-(1.5) become
$${\cal G} = \frac{\beta^2}{2(m+M)}
\int^M_{-m}\Big(\psi^2(\frac{\beta^2\psi^2}{2}-1+A^2)+(\psi')^2
+\frac{1}{r\beta^2}(A'-h)^2\Big)dt,\eqno(1.10)$$
 
$$\psi^{''} = \psi(\beta^2\psi^2+A^2-1),\eqno(1.11)$$
$$A^{''} = r\beta^2\psi^2A,\eqno(1.12)$$
$$\psi'(-m) = \psi'(M) = 0, A'(-m) = A'(M) = h > 0,\eqno(1.13)$$
and 
$$\psi > 0 \quad {\rm in} \quad [-m, M].\eqno(1.14)$$

We now state our existence result:

\medskip
\noindent

{\bf Theorem 2 {\it {\rm For sufficiently small $\beta >0$ there is a }
value $h>1.6831$ and a 
solution $(\psi ,A)$ of (1.11)-(1.14) defined on an interval 
$[-m,M],$ such that the following properties hold:} 
\medskip

{\bf (i)}  $m = m (\beta) > 0, \ M = M(\beta) > 0 \ {\rm and} \
\lim_{\beta\to 0^+}(m,M) = (0,\infty);$

{\bf (ii)}  $\psi' > 0 \ {\rm on} \ (-m, 0), \ \psi'(0) = 0 \ {\rm and} \
\psi(0) = 1;$

{\bf (iii)}  $\psi' < 0 \ {\rm on} \ (0, M);$

{\bf (iv)}  There is an } $ h^0>1.6831$ 
{\rm such that}

$$A'(-m) \to h^0 \quad{\rm as}\quad \beta \to 0^+.\eqno(1.15)$$

\medskip\noindent {\bf Remark:  }\rm Because of (1.8) and (1.9), 
each of the solutions $(\psi ,A)$ found in Theorem 2 
corresponds to a solution $(\tilde{\phi },\tilde {a})$ of (1.2)-(1.5).  Since $
\psi$ has 
a relative maximum at $t=0$, (1.8) implies that 
$\tilde{\phi}'(x_{\max})=0$ where $x_{\max}=\frac d2\big(\frac {m
-M}{m+M}\big).$ It follows from 
(1.9) and properties (i) and (iii) of Theorem 2 that for 
small $\beta >0,\ x_{\max}<0,$ and $\tilde{\phi}'<0\ \ {\rm f}{\rm o}
{\rm r}\,\,\,\,{\rm a}{\rm l}{\rm l}\,x\in (x_{\max},\frac d2).$ 
Therefore $\tilde{\phi}'(0)<0,$ and (1.6) cannot hold.  We conclude 
that $(\tilde{\phi },\tilde {a})$ is an asymmetric solution of (1.2)-(1.5). 

We now state our second result.  Recall that $\cal G$ gives the free 
energy of
a solution.

\medskip
\noindent
{\bf Theorem 3}  {\it Let $(\psi,A)$ be an asymmetric solution found in
Theorem 2.  There exists $\gamma > 0$ such that if $r \in (0, 2 +
\gamma)$ then ${\cal G} < 0$ for sufficiently small $\beta > 0.$}
\medskip
\medskip

\noindent{\bf Asymmetric Minimizers}

\medskip

We now return to the original system (1.1)-(1.5) and show 
that ${\cal G}$ has an asymmetric minimizer.  Recall that 
$k\ge\frac 1{\sqrt {2.01}}$ is fixed.  Also, it follows from (1.7) that if 
$\frac {h_e}k\ge 1.6831$ and $d\gg 1$ then the only symmetric solution of 
(1.2)-(1.5) is the trivial solution $(\tilde{\phi },\tilde {a})=(
0,h_ex),$ also 
known as the ``normal state.'' 
Substitution of this pair into (1.1) shows that ${\cal G}=0.$ 
Next, we conclude from (1.8), (1.9) and the results of 
Theorems 2 and 3 that the problem (1.2)-(1.5) has an 
asymmetric solution for large $d$, that $\tilde {a}'(-\frac d2)\ge 
1.6831\,k,$ 
and that the corresponding energy ${\cal G}$ is negative.  
Therefore, in this parameter range, since ${\cal G}$ is zero for 
the trivial symmetric solution and negative for the 
asymmetric solutions,  a minimizer of ${\cal G}$ must be 
asymmetric.  

\medskip
\noindent
\medskip

{\bf 2.  Proof of Theorem 2.}

Our goal in this section is to show that for small $\beta > 0$ there is
a solution $(\psi, A)$ of the system
$$\psi^{''} = \psi(\beta^2 \psi^2 + A^2-1), \eqno(2.1a)$$
$$A^{''}= r\beta^2\psi^2A,\eqno(2.1b)$$
on an interval $[-m,M],$ where $m > 0$ is small and $M > 0$ is large, and
such that
$$\psi'(-m) = \psi'(M) = 0, \ A'(-m) = A'(M)=h>0,\eqno(2.2a)$$

$$\psi' > 0 \ {\rm on} \ (-m,0), \ \psi'< 0 \ {\rm on} \
(0,M),\eqno(2.2b)$$
$$\psi > 0 \ {\rm on} \ [-m,M],\eqno(2.2c)$$
and
$$\psi(0) = 1, \quad \psi'(0) = 0.\eqno(2.2d)$$

Our method of proof uses a topological shooting 
argument.  For this we begin by analyzing the important 
properties of solutions of the initial value problem (2.1a), 
(2.1b), (2.2d) when $\beta=0$.

In this case 
$A'$ is constant, and we set $A' = h,$ where $h > 0$ is to be determined
later.  Setting $A(0) = A_0,$ also to be determined later, we obtain the
second order linear equation
$$\psi^{^{\prime\prime}}=((A_0+ht)^2-1)\psi .\eqno(2.3)$$
Because of (2.2d) we consider the solution of (2.3) such 
that the $\psi (0)=1,\psi'(0)=0.$ 

\medskip
\noindent
{\bf Lemma 2.1}  {\it Suppose that $-1 \le A_0 \le 0.$  Then there is a
unique $h_0 > 0$ (depending continuously on $A_0)$ such that $\psi > 0,
\psi' < 0$ on $(0, \infty),$ and $\psi(t) \to 0$ as $t \to \infty.$  If $0
< h < h_0$ then $\psi = 0$ before $\psi' = 0,$ while if $h > h_0$ then
$\psi' = 0$ before $\psi = 0.$}

\medskip
\noindent
{\bf Proof:}  We consider the Ricatti equation obtained by setting
$$\rho(s) = \frac{\psi'(s/h)}{\psi(s/h)}.$$
Then $\rho(0) = 0$ and
$$\rho' = \frac{H(s)-\rho^2}{h},\eqno(2.4)$$
where $H(s) = (A_0 + s)^2-1.$  Since $A_0 \in [-1,0], \rho$ initially
decreases, for any $h > 0.$  Further, $\rho' < 0$ as long as $\rho(s)^2
> H(s).$  As long as $\rho' < 0,$ the right side of (2.4) is an
increasing and negative function of $h$ and $\rho.$  Suppose, for some
first $s_0>0,$ that $\rho'(s_0)=0.$  Then $H(s_0)=\rho (s_0)^2$ so that 
$A(s_0)>1.$ Therefore, $\rho^{\prime\prime}(s_0)=\frac {2A(s_0)}h>0.$  
For
$s>s_0$ it follows from the equation for $\rho''$ that $\rho^{^{\prime\prime}}>0$ so that $\rho$ increases until
$\rho = 0;$ i.e. $\psi' = 0.$  Also, if $\rho'$ becomes positive for
some $h_1,$ because $\rho$ crosses the curve $\rho^2 = H(s),$ then the same must
happen for any $h>h_1.$ To see that there are values of$ $ $h$ 
such that $\rho'$ becomes positive we refer to (2.3).  From 
that equation and our assumption that $\psi (0)=1,$ $\psi'(0)=0$ 
we easily see that for large $h,$  $\psi'$ becomes positive 
before $\psi =0.$ This implies that $\rho ,$ and hence $\rho'$,  must 
become positive for large $h$.  It then follows from continuity that
 the set of values 
of $h>0$ such that this happens is open. 

For small $h>0,$ on the other hand, we see that $\rho$ 
decreases to below the curve $\rho =-s.$ For example, if 
$-1<A_0<0$ and $0<h<1-A_0^2$ , then $\rho'(0)<-1,$ so 
immediately $\rho$ decreases below $-s.$ If $A_0=-1,$ then 
$\rho (0)=\rho'(0)=0.$  However, $\rho^{\prime\prime}(s)<\frac {2
s-2}h$ and this integrates 
to show that for small $h,$ $\rho (2)<-2$ and $\rho'<0$ over $(0,
2].$  
It  follows 
from (2.4), and our assumption that $A_0\in [-1,0],$ that if $\rho'$ 
continues to decrease until $\rho\to -\infty$ at some finite $s_1
.$ 
That is, $\psi (s_1)=0.$ It is clear that the set of $h's$ such 
that $\rho (s)<-s$ for some $s$ is an open subset of the 
interval $0<h<\infty .$ Similarly, as we pointed out above, 
the set of $h's$ such that $\rho (s)^2<H(s)$ for some $s$ is also 
open.  

   Moreover, if $\rho$ ever falls below $-s$, then we see from the
equation for $\rho''$ that thereafter, $\rho'<-1$, and $\rho$ decreases 
monotonically to $-\infty$.  On the other had, if $\rho'$ is ever positive, 
then we saw that $\rho$ becomes positive, and because of (2.4) it must
remain positive.

 Hence there is at least one $h_0$ such that on $0\le s<\infty$, $\rho'\le 0$
and $0>\rho>-s$. These bounds imply that the solution exists
 on the entire interval $[0,\infty ).$ Further 
properties of this solution are given in the following 
result. 

\medskip
\noindent
{\bf Lemma 2.2}  {\it For such an $h_0, \rho' < 0, \  -s < \rho(s)$ on
$(0,\infty), \rho(s) < -\sqrt{H(s)}$ if $H(s) \ge 0,$ and $\rho +
\sqrt{H(s)} \to 0$ as $s \to \infty.$}

\medskip
\noindent
{\bf Proof:}  The inequalities have already been proved.  To see the
limiting behavior of $\rho(s),$ we first note that for $h = h_0,
\rho'(s)$ is bounded.  If $\rho'$ is unbounded, then it must get
arbitrarily large and negative, but then the equation for $\rho^{''}$
shows that $\rho'$ remains large and negative and $\rho(s) < -s$ for
some $s.$    

Writing $\rho'=\frac {(\sqrt {H}-\rho )(\sqrt {H}+\rho )}h$ and noting that the first of 
these factors is unbounded, we see that $\sqrt {H(s)}+\rho (s)$ tends 
to zero, which implies Lemma 2.2.  

To prove uniqueness we suppose that there is a second positive value of $h,h_1<h_0$ 
for which this behavior occurs.  Let $\rho_1,\rho_0$ denote the 
corresponding solutions.  Then $\rho_1(0)=\rho_0(0)=0,$ and it is 
easily shown that 
\[\frac d{ds}(\rho_1-\rho_0)<0\]
for all $s>0.$ Thus $\rho_1-\rho_0$ could not approach zero as 
$s\to\infty .$ However, Lemma 2.2 implies that 
$\rho_1-\rho_0=(\rho_1+\sqrt {H})-(\rho_0+\sqrt {H})\to 0$ as $s\to
\infty ,$ a contradiction.  
Therefore $h_0$ is unique.  

To complete the proof of Lemma 2.1 we must show that $h_0$ depends
continuously on $A_0.$  This follows from the uniqueness of $h$ by a
standard argument.

Now let $A_0=-1,$ and define
\[h^0=\lim_{\beta\to 0}h_0(\beta ).\]
We will find the desired solutions near the point 
$(A_0,h)=(-1,h^0).$

\medskip
\noindent
{\bf Lemma 2.3}  $h^0>1.6831.$

\medskip
\noindent
{\bf Proof:}  Let
$$\hat{\rho}(t) = \frac{\psi'(t)}{\psi(t)},$$
so that 
$$\hat{\rho}' = H(ht) - \hat{\rho}^2.\eqno(2.5)$$

Here, because $A_0=-1,$ $H(s)=$ $-2s+s^2,$ where $s=ht$ as before.  
We use the following result about a solution $\hat{\rho}$ of (2.5) such
that $\hat{\rho }(0)=0.$  If $\hat{\rho}^{^{\prime\prime\prime}}<
0$ for some $t_1$, with
$\hat{\rho },\hat{\rho}',\hat{\rho}^{^{\prime\prime}}<0$ on $(0,t_1
)$ then $\hat{\rho}$ decreases
monotonically to $-\infty.$  To see this, compute the equation satisfied
by $\hat{\rho}^{''''}.$  That is,
$$\hat{\rho}^{''''} + 2 \hat{\rho}\hat{\rho}^{'''} = -6
\hat{\rho}'\hat{\rho}^{''}.$$
  
If there were a first $y>t_1$ where $\hat{\rho}^{^{\prime\prime\prime}}
(y)=0$ then
$\hat{\rho}^{''''}(y) \ge 0.$  However, the last equation gives
$\hat{\rho}^{''''} (y) < 0$ since the definition of $y$ implies that
$\hat{\rho}'\hat{\rho}^{^{\prime\prime}}$ is positive and increasing on $(t_1
,y).$  From these derivative 
properties it follows that $\hat{\rho}$ decreases below $-s,$ and 
hence $\hat{\rho }(s)$$\to -\infty$ at a finite value of $s.$ Next, we
define a sequence $\{\hat{\rho}_N\}$ of functions, all defined on
$[0,2]$ as follows:
\[\hat{\rho}_0(t)=\int^t_0(h^2r^2-2hr)dr\quad ,\quad\hat{\rho}_{N
+1}(t)=\hat{\rho}_0(t)-\int^t_0\hat{\rho}_N(r)^2dr\quad for\quad 
N\ge 1.$$
From this definition it is evident that $\{\hat{\rho}_N\}$ forms a 
decreasing sequence of functions defined on [0,2], for all 
$N\ge 1.$ Furthermore, it is easily shown that our solution 
$\hat{\rho }(t)$ satisfies $\hat{\rho}^{^{\prime\prime\prime}}<\hat{
\rho}_i^{^{\prime\prime\prime}}$ on [0,2], for all $i\ge 1.$ Setting 
$h=\frac {16831}{1000},$ we used the computer algebra program
 Maple to compute $\hat{\rho}_6,$ a polynomial of 
degree 225. All calculations are with integers and rational 
numbers so that there are no roundoff errors. 
From $\hat{\rho}_6$ and its first two derivatives we 
can compute $\hat{\rho}_7^{\prime\prime\prime},$ without having to compute $
\hat{\rho}_7.$ We find 
that $\hat{\rho}_7^{\prime\prime\prime}(6/5)<0$, and that on [0,6/5],   $
\hat{\rho },\hat{\rho}',\hat{\rho}^{^{\prime\prime}}$ 
are all negative.  This completes the proof of the 
lemma.  

\medskip

Our solution $(\psi ,A)$ to (2.1a)-(2.2d) is obtained by perturbing 
$A(0)$ from
$-1$, keeping $\beta =0,$ and then letting $\beta$ be positive.
  Our argument is by ``shooting'', rather than by use of bifurcation theory.
 
 Thus,
we will assume that
$$A(0) = -1 + \epsilon$$
for small $\epsilon \ge 0,$ and we let $h_0 = h_0(\epsilon)$ denote the
value of $h$ found in Lemma 2.1.

In constructing solutions to (2.1)-(2.2), in order to get something
meaningful at $\beta = 0,$ we replace the boundary conditions on $A$ in
(2.2a) with
$$\int^M_{-m} \psi^2(t)A(t)dt = 0.$$

\medskip
\noindent
{\bf Lemma 2.4}  {\it  Suppose that $\epsilon = 0,$ (i.e. $A_0 = -1)$ and let $\psi_0$ be
the solution found in Lemma 2.1 where $h = h^0 = h_0(0).$  Then
$$\int^\infty_0 \psi_0(t)^2(-1+h^0t)dt = 0.$$
}

\medskip\noindent {\bf Proof:}  It is easily seen that $\psi_0(t)
\to 0$ 
exponentially fast as $t\to\infty .$ Thus, the integral in the 
lemma converges and $\lim_{t\to\infty}(-1+h^0t)^2\psi_0(t)^2=0.$ We set 
$\psi =\psi_0$ in (2.3), multiply by $\psi_0',$ and integrate by parts 
to obtain the result.

  From Lemma 2.1 and the definition of $h_0(\epsilon )$ we see 
that for each $h>h_0(\epsilon ),$ there is a first $t=t_h>0$ such 
that $\psi'(t_h)=0,\psi'<0$ and $\psi >0$ on $(0,t_h),$ with $\psi 
(t_h)>0.$ 
 By the implicit 
function theorem, $t_h$ is continuous in $h$ on $(h_0(\epsilon ),
\infty )$ since 
$\psi^{^{\prime\prime}}(t_h)>0.$ (If $\psi^{^{\prime\prime}}(t_h)
=0$ then $t_h=\frac 2h$ and $\psi^{^{\prime\prime\prime}}(t_h)>0,$ a 
contradiction.) Further, $t_h\to\infty$ as $h\to h_0(\epsilon)$ from above.

Now consider small $\epsilon > 0,$ set $h = h_0(\epsilon),$ and
compute $\psi^{''}(0)$ and $\psi^{'''}(0).$  We find that $\psi{''}(0) =
-2\epsilon + \epsilon^2$ and $\psi^{'''}(0) =
2(-1+\epsilon)h_0(\epsilon).$  From this and the fact that
$h_0(\epsilon) \to h^0 > 0 \ {\rm as} \ \epsilon\to 0^+,$ we see
that for small $\epsilon, \psi'(-m) = 0$ for some $m = m_0(\epsilon) >
0.$  Furthermore,

$$-m = \frac{-2\epsilon}{h_0(\epsilon)}+0(\epsilon^2).\eqno(2.6)$$
Here, $O(\epsilon^2) < L \epsilon^2$ for some $L$ independent of
$\epsilon.$  We now compute
$$I(\epsilon) = \int^\infty_{-m} \psi^2  A  ds$$
where $A(s) = -1 + \epsilon + h_0(\epsilon)s.$  Again, multiplying (2.3)
by $\psi'$ and integrating by parts gives

$$0 = -\psi(-m)^2A(-m)^2+ \psi(-m)^2-2h_0\int^\infty_{-m}\psi(s)^2 A(s)
ds.$$

Using (2.6) and the fact that $\psi (-m)=1+O(\epsilon ),$ we find 
that $I(\epsilon )=-\frac {\epsilon}{h_0}+O(\epsilon^2)$ as $\epsilon
\to 0^{+}.$ Thus, we have shown 
that with $h=h_0(\epsilon )$ for small $\epsilon ,\,\,\,I(\epsilon 
)<0$.  Fix $\epsilon >0$ 
small enough that this inequality holds for $h=h_0(\epsilon )$.  Then for 
$h-h_0(\epsilon )$ positive but sufficiently small, $-m=-m(\epsilon 
,h)$ will 
still be defined as the largest negative zero of $\psi',$ $A$ is 
positive on $(t_h,\infty ),$ and therefore 
\[\int^{t_h}_{-m}\psi^2Ads<0.\]

Our goal now is to find an $h$ and $\epsilon$ such that $-m$ and $t_h$
are defined as the negative and positive zeros of $\psi'$ closest to $t
= 0,$ with $m$ small and $t_h$ large, and such that
\[I=\int^{t_h}_{-m}\psi^2Ads>0.\]
We will see that it is not necessary to have the same $m$ 
and $t_h$ as before.  Starting with $\epsilon =0$ and $h=h_0(\epsilon 
),$ we 
now raise $h$, instead of $\epsilon .$ Thus, for small $h-h^0>0$ our 
solution satisfies $A(0)=-1,\,\,\,A'(0)=h>h^0=h_0(0),\,\,\,\psi'(
t_h)=0$ 
for some large $t_h.$ In this case, we multiply (2.3) by $\psi'$ 
and integrate from $0$ to $t_h,$ where $t_h$ is the first 
positive zero of $\psi'.$ Using integration by parts once 
again, we find that $I>0.$ We then keep $h$ fixed and raise 
$\epsilon$ slightly, whereupon both $m=m(\epsilon ,h)$ and $t_h$ are defined, 
still with $I>0.$ We summarize these results in the 
following lemma.  

\medskip
\noindent
{\bf Lemma 2.5}  {\it For each sufficiently small $\epsilon > 0$ there
is an $h_1(\epsilon) > h_0(\epsilon)$ such that if $h_0(\epsilon) < h <
h_1(\epsilon)$, then the solution of (2.3) with $A_0 = -1 + \epsilon,
\psi(0) = 1,$ and $\psi'(0) = 0$ is decreasing on an interval $[0,M]$,
increasing on an interval $[-m,0],$ and $\psi'(-m) = \psi'(M) = 0,$ and

$$I(\epsilon,h) = \int^M_{-m}\psi^2  A  ds < 0.$$
Furthermore, $\psi > 0$ on $[-m,M]$ and $h_1(\epsilon) \to h_0(0) = h^0$ as $\epsilon \to 0^+.$
On the other hand, for each $h > h_0(0)$ sufficiently close to $h_0(0),$
there is an interval $(0, \epsilon_1(h))$ of $\epsilon's$ such that
$\psi$ has the same behavior, but $I(\epsilon,h) > 0.$  As $\epsilon$ and
$h-h_0(0)$ tend to zero, $m \to 0$ and $M \to \infty.$}

\medskip
\noindent
{\bf Corollary 2.6}  {\it  For $\epsilon > 0$ sufficiently small, there
is an $h_2 = h_2(\epsilon)$ such that if $\psi$ is the solution of (2.3)
with $A_0 = -1 + \epsilon$ and $(\psi(0), \psi'(0)) = (1,0)$, then there
are values $M > m > 0$ such that}

$$\psi'(-m)=\psi'(M)=0,\psi'>0\ {\rm o}{\rm n}\ (-m,0),\psi'<0\ {\rm o}
{\rm n}\ (0,M),\eqno(2.7a)$$
$$\psi^{^{\prime\prime}}(-m)\ne 0,\psi^{^{\prime\prime}}(M)\ne 0.\eqno
(2.7b)$$

{\it Further, $I(\epsilon, h_1(\epsilon)) < 0$ and $I(\epsilon,
h_2(\epsilon)) > 0.$  Finally, $h_2(\epsilon) \to h_0(0) = h^0$ as
$\epsilon \to 0.$}

\medskip

With $\epsilon >0$ sufficiently small so that Lemma 2.5 and 
Corollary 2.6 hold, we now raise $\beta$ and consider solutions 
of (2.1) satisfying (2.2d).  Since $\psi^{^{\prime\prime}}\ne 0$ at the zeros of $
\psi'$ 
when $\beta =0$, it follows from the Implicit Function 
Theorem that the zeros $-m<0$ and $M>0$ of $\psi'$ persist 
as continuous functions of $\beta$ and the values of $A(0)$ and 
$A'(0)$. For each sufficiently small 
$\epsilon >0$ there exist functions $
h_1(\epsilon )$ and $h_2(\epsilon )$ 
independent of $\beta$, such that for sufficiently small $\beta$
(depending on $\epsilon$) the solution $(\psi ,A)$ of (2.1) and 
(2.2d) with 
$$A(0)=-1+\epsilon ,\,\,\,A'(0)=h_1(\epsilon ),\eqno(2.8)$$
satisfies (2.7)  on an interval $[-m,M],$ and 
$$\int_{-m}^M\psi^2Adt<0.$$
Also, the solution of (2.1), (2.2d) with $A(0)=-1+\epsilon ,$ 
$A'(0)=h_2(\epsilon )$  satisfies 
\[\int_{-m}^M\psi^2Adt>0.\]
Furthermore, $\beta\to 0$ as $\epsilon\to 0,$ and
\[(m,M)\to (0,\infty ),\,\,\,\,\,(h_1(\epsilon ),h_2(\epsilon ))\to 
(h^0,h^0).\]
In
addition, it follows from (2.1b) that $A^{^{\prime\prime}}$ is uniformly
 bounded on $[-m,0]$,
and therefore $A'(-m)=h\to h^0$ as $\epsilon\to 0$ and $\beta\to 0^{+}.$  It then follows 
from continuity that for given fixed small $\epsilon$, and
sufficiently small $\beta\ge 0$, there is an $h\in (h_1(\epsilon ),h_2(\epsilon 
))$ such that 
the solution $(\psi_{\epsilon,\beta} ,A_{\epsilon,\beta})$
of (2.1),  (2.2d),  and (2.8)  with 
$A'(-m)=h$ also satisfies $\int_{-m}^M\psi^2Adt=0$, so $A'(M)=h$. 
The conclusion of Theorem 2 now follows from the
transformation (1.8)-(1.9).
\medskip
\medskip

In the next section we turn to the proof of Theorem 3.  For this we will
use the following result.  Let
$${\cal E}(\psi,A) = \int^M_{-m}F(\psi,A)(t)dt$$
where

$$F(\psi,A)(t) =
r\Big(\int^M_t\psi(s)^2A(s)ds\Big)^2-\frac{\psi(t)^4}{2}.$$

\medskip
\medskip
\noindent
{\bf Lemma 2.7}.  {\it{With $(\psi ,A_{})=(\psi_{\epsilon ,\beta}
,A_{\epsilon ,\beta})$ chosen as in the proof of 
Theorem 2, 
we have
\[\lim_{\epsilon\to 0}\{\lim_{\beta\to 0}{\cal E}(\psi ,A)\}={\cal E}
(\psi_0,A^0)\]
where $A^0(s) = -1 + h^0s.$}
}
\medskip
\medskip
\noindent

{\bf Proof:}  Our proof of Theorem 2 shows that
$$\lim_{\epsilon\to 0}\{\lim_{\beta \to 0}(\psi_{\epsilon,\beta}(s),
A_{\epsilon,\beta}(s))\}=(\psi_0(s),A^0(s))$$
uniformly on compact intervals. 
 Let $\delta >
0$ be given.  We can choose $K_1 > 0$ such that
$$r\psi(K_1)^4<\delta\, {\rm  }\int^\infty_{K_1}\big|F(\psi_0,A^0)(t)\big|dt < \delta,\eqno(2.9)$$
and
$$A^0(s)^2-2 > \frac{A^0(s)^2}{2},\eqno(2.10)$$
for $s \ge K_1.$
Then for sufficiently small $\epsilon>0$ 
we choose $\beta_1 =\beta_1(\epsilon) 
> 0$ such that for $0 < \beta < \beta_1,$
$$\Big|\int^{K_1}_0(F(\psi,A)(t) - F(\psi_0,A^0)(t))dt\Big| < \delta.\eqno(2.11)$$

Further, we can insure that for $0 < \beta < \beta_1,$ (2.10) also
holds on $[K_1,M]$ with $A$ substituted for $A_0.$

We consider two cases:

$$\frac {\psi_0'(K_1)}{\psi_0(K_1)}>-1\eqno(i)$$
and 
$$\frac {\psi_0'(K_1)}{\psi_0(K_1)}\le -1.\eqno(ii)$$
Let $\rho =\frac {\psi'}{\psi},$ so that, from (2.1a),
$$\rho'(s)\ge A(s)^2-1-\rho^2.$$
In case (i) we have $\rho'\ge\frac {A(s)^2}2$ on $[K_1,M]$ as long as $
-1\le \rho<0,$ 
so that $\rho =0$ (and hence $\psi'=0$)  before 
$$\int_{K_1}^t\frac {A(s)^2}2ds=1.$$
In other words,  
$$\int_{K_1}^M\frac {A(s)^2}2 ds\le 1.\eqno(2.12)$$
Thus, (2.10) implies that if $M/ge t \ge K_1,$ then $\frac{A(s)}(2)/ge1$ on
$[t,M]$, and hence
\[\int_t^M\psi (s)^2A(s)ds\le\psi (K_1)^2\int_t^M\frac {A(s)^2}2d
s\le\psi (K_1)^2.\]
Since (2.10) and (2.12)  also imply that $M-K_1<1,$ 
it follows from (2.12) and (2.9) that in case (i), 
$$\int_{K_1}^MF(\psi ,A)(t)dt<\delta\eqno(2.13)$$
for sufficiently small $\epsilon$ and $\beta$.
\medskip
 In case (ii),  since $\psi'(M)=0,$  there must be a 
$T\in [K_1,M)$ such that $\rho\le -1$ on $[K_1,T]$ and $\rho\ge -1$ on $
[T,M].$
\medskip
 First consider $\int_{K_1^{^{}}}^TF(\psi ,A)(t)dt$.
On $[K_1,T]$ we have $\psi'\le -\psi ,$ so that 

$$\psi (t)\le\psi (K_1)e^{K_1-t}.\eqno(2.14)$$
We now estimate the term 
$$\int_t^M\psi (s)^2A(s)ds$$
in $F(\psi ,A)$ for $K_1\le t\le T.$ For this we use: 
\medskip

\noindent
{\bf Lemma 2.9} $\lim_{t\to\infty}\psi_0(t)A^0(t)=0.$

\medskip
\noindent

{\bf Proof:}  
In our estimates of $\psi_0$ we saw 
that $\frac {\psi_0'}{\psi_0}+H_0(t)\to 0.$ Since $A^0$ grows linearly, the result 
follows.
\medskip

We now continue with the proof of Lemma 2.7.  We choose $K_1$ so that in addition to the earlier 
constraints,  $\psi_0(K_1)A^0(K_1)<1.$ For small $\epsilon$ and 
 $\beta ,$ this 
inequality will also be satisfied by $(\psi ,A).$ If 
$t\le T,$ then 
$$\int_t^M\psi (s)^2A(s)ds=\int_t^T\psi (s)^2A(s)ds+\int_T^M\psi 
(s)^2A(s)ds$$
$$\le\int_t^T\psi (s)^2A(s)ds+\psi (T)^2\int_T^MA(s)ds.$$
The argument used earlier to get (2.12)
 shows that $\int_T^MA(s)ds\le 1,$ where we use (2.10) with $A$ 
substituted for $A^0$.
From an integration of $\frac {\psi'}{\psi}\le -1$ we obtain  
\[\int_t^T\psi (s)^2A(s)ds\le\psi (t)(\max_{t\le s\le T}\psi (s)A
(s))\int_t^Te^{t-s}ds.\]
In $[K_1,T],$ $\psi'\le-\psi ,$ so $\frac d{ds}(\psi (s)A(s))$$\le -\psi 
(s)A(s)+\psi (s)A'(s).$
But $A'(s)\le A'(M)=h\le 2h^0$ for small $\epsilon$ and  $\beta ,$ so $
\psi (s)A(s)$ is 
decreasing in $[K_1,T].$ Hence,  
$\int_t^T\psi (s)^2A(s)ds\le\psi (t)\le\psi (K_1)e^{K_1-t}$.  Using this
 we 
find that  
on $[K_1,T],$
$$F(\psi ,A)\le r\psi (K_1)^2e^{2(K_1-t)}.$$
Thus, we can further restrict $K_1$ to insure that for 

small positive $\epsilon$ and $\beta ,$ 

$$\int_{K_1}^TF(\psi ,A)(t)dt<\delta .$$
Finally, we consider $\int_T^MF(\psi ,A)(t)dt$.  As in case (i), we 
have 
$$\int_T^M\frac {A(s)^2}2ds\le 1.$$
Hence we find that 
$$\int_t^M\psi (s)^2A(s)ds\le\psi (T)^2\le\psi (K_1)^2$$
and 
$$M-T\le 1.$$
We then have, (without further change in $K_1)$ that 
$$\int_T^MF(\psi ,A)(t)dt<\delta .$$
Since $\delta$ was arbitrary,  this completes the proof of 
Lemma 2.7.

\medskip
\medskip

\noindent{\bf 3.  Proof of Theorem 3}
\medskip

In order to prove Theorem 3 we need to show that the asymmetric
solutions found in Theorem 2 have the additional property that their
corresponding energy $\cal G$ is negative if $\epsilon >0$ and 
$\beta > 0$ are small
enough.  Thus, we define

$$Q = \frac{2(m+M)}{\beta^4} {\cal G}.\eqno(3.1)$$
Then, by (3.1) and (1.10), $Q$ is given by

$$Q = \frac{1}{\beta^2}
\int^M_{-m}\Big(\psi^2\Big(\frac{\beta^2\psi^2}{2} - 1 + A^2\Big) +
(\psi')^2 + \frac{1}{r\beta^2}(A'-h)^2\Big)dt.\eqno(3.2)$$
Therefore, if we show that $Q < 0$ for small $\epsilon>0$ and 
$\beta > 0$ then
$\cal G$ also is negative and Theorem 3 is proved.  We need to simplify
$Q.$  For this we use (1.11) and conclude that

$$\psi^2\big(\frac{\beta^2\psi^2}{2}-1+A^2\big) + (\psi')^2 = \psi
\psi^{''} + (\psi')^2 - \frac{\beta^2\psi^4}{2}.\eqno(3.3)$$
Then substitution of (3.3) ino (3.2), together with the observation that
$(\psi \psi')' = \psi \psi^{''} + (\psi')^2,$ reduces $Q$ to
$$Q = \frac{1}{\beta^2} \int^M_{-m} ((\psi \psi')' -
\frac{\beta^2\psi^4}{2} + \frac{1}{r\beta^2}(A'-h)^2)dt.\eqno(3.4)$$
Since $\psi'(-m) = \psi'(M) = 0,$ (3.4) further reduces to
$$Q =
\frac{1}{\beta^2}\int^M_{-m}\Big(\frac{1}{r\beta^2}(A'-h)^2
-\frac{\beta^2\psi^4}{2}\Big)dt.\eqno(3.5)$$
  
Next, it follows from an integration of (1.12) that
$$h-A' = r\beta^2 \int^M_t\psi^2 A ds.\eqno(3.6)$$
Finally, we substitute (3.6) into (3.5) and arrive at
$$Q = \int^M_{-m}\Big(r\Big(\int^M_t \psi^2
Ads\Big)^2-\frac{\psi^4}{2}\Big) dt.\eqno(3.7)$$
In view of Lemma 2.7 we conclude from (3.7) that
$$\lim_{\epsilon \to 0} \lim_{\beta\to 0} Q = \int^\infty_0\Big
(r\Big(\int^\infty_t
\psi^2_0A^0 ds\Big)^2 - \frac{\psi^4_0}{2}\Big)dt\eqno(3.8)$$
where $A^0 = -1 + h^0s.$

In the next section we prove
\medskip

\noindent{\bf Lemma 3.1.}  There is a value $\gamma > 0$ such that
\medskip

$$\int^\infty_0(r\Big(\int^\infty_t \psi^2_0 A^0ds\Big)^2 -
\frac{\psi_0^4}{2}) dt < 0 \eqno(3.9)$$
for all $r \in (0, 2 + \gamma).$

From Lemma 3.1 we observe that if $r \in (0, 2 + \gamma)$ then
$Q < 0$ for small $\epsilon>0$ and $\beta > 0.$  Therefore ${\cal G}$ is also
negative for small $\beta > 0$ and Theorem 3 is proved.

\medskip

\noindent{\bf 4.  Proof of Lemma 3.1}
\medskip

For the proof of Lemma 3.1 we recall that $\psi_0$
satisfies 
$$\psi^{''}_0 = \psi_0(-2h^0t + (h^0)^2t^2),\eqno(4.1)$$
$$\psi_0(0) = 1, \psi'_0(0) = 0\eqno(4.2)$$
where  $h^0 > 0$ is the unique positive value for which
$\psi_0$ satisfies $\psi'_0 < 0$ for all $t > 0,$ and
$$\lim_{t\to\infty}(\psi_0(t), \psi_0'(t)) = (0,0).\eqno(4.3)$$
The existence and uniqueness of $h^0$ was proved in Lemma 2.1. 
Thus, our main objective is to prove
\medskip

{\bf Lemma 4.1}  {\it There is a value $\gamma > 0$ such that}
\hspace{\parindent}

$$\int^\infty_0\Big(r\Big(\int^\infty_t \psi_0^2(-1+h^0s)ds\Big)^2-\frac 12
\psi^4_0(t)\Big) dt < 0 \eqno(4.4)$$
for all $r\in(0,2+\gamma).$

The proof of Lemma 4.1 relies on an auxiliary result which we now
establish. 

Define the Ricatti variable $q = \frac{\psi'(t)}{\psi(t)}.$  Then $q$
satisfies $q(0) = 0$ and

$$q' + q^2 + 2h^0t - (h^0)^2t^2 = 0.\eqno(4.5)$$

$$q(0) = q'(0) = 0\eqno(4.6)$$

\noindent{\bf Lemma 4.2}  {\it It follows from (4.5) and the definition of $q$
that}

\medskip

$$q<0\quad {\rm a}{\rm n}{\rm d}\quad q'<0\quad {\rm f}{\rm o}{\rm r}\,\,
{\rm a}{\rm l}{\rm l}\,\,\,\,t>0,\eqno(4.7)$$

$$q\le -\sqrt {(h^0)^2t^2-2h^0t}\quad {\rm f}{\rm o}{\rm r}\,\,{\rm a}
{\rm l}{\rm l}\,\,\,t\ge\frac 2{h^0},\eqno(4.8)$$

$$q'>-h^0\quad {\rm f}{\rm o}{\rm r}\,\,{\rm a}{\rm l}{\rm l}\,\,\,
t\ge 0.\eqno(4.9)$$

\medskip
\noindent
{\bf Proof:}  { Setting $h = h^0$ and $s = t\, h^0,$ we
observe that (4.7) and (4.8) follow immediately from Lemma 2.2.  It remains
to prove (4.9).  It follows from (4.6) that
$q'(0) = 0.$  Thus $q' > - h^0$ on an interval $[0,\eta)$ for small
$\eta > 0.$  Suppose that (4.9) is false.  Then there is a first
$\hat{t} > 0$ for which $q'(\hat{t}) = -h^0,$ and $q^{''}(\hat{t})
\le 0.$  Two differentiations of (4.5) lead to

$$q^{^{\prime\prime\prime}}+2qq^{^{\prime\prime}}=2((h^0)^2-(q')^
2).\eqno(4.10)$$
One solution of (4.10) is $q' \equiv - h^0$ for all $t$.   Thus,
if $q^{''}(\hat{t}) = 0$ then uniqueness of solutions implies that $q' =
-h^0$ for all $t \ge 0,$ and in particular, $q'(0) = -h^0$,
contradicting the fact that $q'(0) = 0.$  Therefore it must be the case
that
$$q'(\hat{t}) = h^0 \quad {\rm and} \quad q^{''}(\hat{t}) <
0.\eqno(4.11)$$
It then follows from (4.10) and (4.11) that $q^{''}(t) < 0$ for all $t >
\hat{t}$ so that 
$$\lim_{t\to\infty} q'(t) < - h^0.\eqno(4.12)$$
We conclude from (4.12) that
$$\lim_{t\to\infty} q(t) + h^0 t < 0.\eqno(4.13)$$
However, it follows from Lemma 2.2 that $\lim_{t\to\infty} q(t) +
h^0 t = 0,$ contradicting (4.13).  Thus, it must be the case that
$q' > - h^0$ for all \ $t > 0$ and the proof is complete.}

We now return to the proof of Lemma 4.1.  First, we conclude from
(4.7) and (4.8), and the properties that $0 < \psi_0 < 1$ and
$\psi_0' < 0$ for all $t > 0,$ that the integral
$\int^\infty_0 \psi_0^4(t) dt$ is well defined and
positive.  The same reasoning shows that
$$J = \int^\infty_0\Big(\int^\infty_t \psi^2_0(s)(-1 +
h^0s)ds\Big)^2dt\eqno(4.14)$$
is well defined and positive.  It
remains to prove (4.4).  For this we begin by
estimating the integral
$$H = \int^\infty_t \psi_0^2(-1 + h^0s)ds.\eqno(4.15)$$
An integration by parts reduces (4.15) to
$$H = -\frac{\psi_0^2}{2h^0}(-1 +
h^0t)^2-\frac{1}{h^0}\int^\infty_t
\psi_0'\psi_0(-1+ h^0s)^2ds\eqno(4.16)$$
 
It follows from (4.1) that $\psi_0'\psi_0^{''} + \psi_0'\psi_0 = \psi_0'\psi_0(-1 +
h^{0} t)^2$ and therefore (4.16) becomes
$$H = -\frac{\psi_0^2}{2 h^0}(-1+ h^0t)^2+\frac{\psi_0^2(t)}{2h^0} +
\frac{1}{2h^0}(\psi'_0)^2.\eqno(4.17)$$
 
Recall that $\psi' = q\psi.$  Then (4.6) and (4.17) imply that
$$H = -\frac{\psi_0^2q'}{2h^0}.\eqno(4.18)$$
It follows from (4.18) and (4.14) that
$$J = \int^\infty_0H^2(t) dt \le \int^\infty_0
\frac{\psi_0^4(t)}{4(h^0)^2}(q')^2dt.\eqno(4.19)$$
Because Lemma 4.2 gives $-h^0 < q' < 0$ for all $t > 0,$ (4.19)
further reduces to
$$J < \frac 14 \int^\infty_0 \psi_0^4(t) dt.\eqno(4.20)$$
Finally, it follows from (4.4), (4.14) and (4.20) that there exists
$\gamma > 0$ such that

$$\int^\infty_0\Big(r\Big(\int^\infty_t \psi_0^2(-1+h^0s)ds\Big)^2 -
\frac{\psi_0^4}{2}\Big)dt < 0$$
for all $r\in [0,2+\gamma ).$ This completes the proof of Lemma 4.1.

\medskip
 
\noindent {\bf Appendix:}  In [9], when studying symmetric 
solutions, we used a different scaling from that used in 
this paper.  That is, we rescaled (1.2)-(1.5) by setting 
\[K=\frac {k^2d^2}4,h=\frac {h_ed}2,r=\frac 1{k^2},y=\frac {2x}d,\]
and defined new dependent variables $\phi$ and $a$ by
\[\phi (y)=\tilde{\phi }(x),a(y)=\tilde {a}(x).\]
This transforms the problem (1.2)-(1.5) into

\medskip

$$\phi^{\prime\prime}=K\phi (\phi^2+a^2-1),$$
$$a^{\prime\prime}=rK\phi^2a,$$
\[\frac {d\phi}{dy}(\pm 1)=0,\ \frac {da}{dy}(\pm 1)=h,\]
$$\phi > 0 \ {\rm on} \ [-1,1].$$
Since we are considering symmetric solutions, we consider the initial values 
$$\phi (0)=\beta ,\,\,\,\phi'(0)=0,\,\,\,a(0)=0,\,\,\,a'(0)=\alpha 
.$$
In [11], Kwong proved that for each $\beta \in (0,1)$ there exists a
unique $\alpha = \alpha(\beta) > 0,$ continuously dependent and
decreasing in $\beta,$ such that $\phi'(1, \beta, \alpha(\beta)) = 0.$ 
He then set $h(\beta) = a'(1, \beta, \alpha(\beta))$ and proved that $h$
is continuous, $h(0) > 0$ and $h(1) = 0.$  Thus, to obtain the upper
bound $h_{sym} \le \sqrt 3$ given in [9] we found that it was sufficient
to fix $r \in (0, 2.01]$ and estimate
$$h_{sym} = \overline{\lim}_{K \to \infty}\Big(\sup_{0 \le \beta \le 1}
\frac{h(\beta)}{\sqrt K}\Big).$$
In [9], our estimate for $h_{sym}$ was obtained by carefully analyzing the
behavior of $(\phi,a)$ for each $\beta \in (0,1).$  We considered three
intervals of $\beta,$ namely $(0,.1], (.1, 1-\frac{1}{\sqrt K}]$ and
$(1-\frac{1}{\sqrt K}, 1).$
   We made repeated use of the ``energy'' function

$$\frac {\phi^{\prime 2}}K+\frac {a^{\prime 2}}{rK}-\frac {\phi^4}
2+\phi^2-a^2\phi^2,$$
which is constant for solutions of the system.
   We defined the function 
\[Q(y)=\beta^2-\frac {\beta^4}2+\frac {\phi (y)^4}2-\phi (y)^2+a^
2\phi (y)^2,\]
and found the point $y_0$ where $Q(y_0)=1$ to be  
important.   \medskip\ The difficult part of the estimate 
required us to consider a small  interval of values of 
$\phi (y_0)$, say $I_1<\phi (y_0)<I_2$.   A priori we know only
that $0<\phi (y_0)<1$, and to get better estimates
 we had to  subdivide $(0,1)$ into nine 
small subintervals $[I_1,I_2]$. 

  \medskip\ In this paper, to get the improved 
estimates required to show that $h_{sym}\le 1.68$  simply 
requires that we use more subintervals.  Thirty-four subintervals  
suffice.  They are:  [0,.407], [.407,.56],[.56,.6], then thirty 
intervals from .6 to .9 in steps of .01, and finally [.9,1.0].  

\medskip

With this change, the proof in [9] gives the required 
upper bound for $h_{sym}$, and because of our rescaling this immediately
leads to $h_e^{sym}\le 1.68k$ for $k\ge\frac 1{\sqrt {2.01}}$.  As we mentioned  
in [9], our estimate for $h_{sym}$  is much easier for small $r$.

\end{document}